\documentclass[preprint,12pt]{elsarticle}
\usepackage{amsfonts}
\usepackage{graphicx}
\usepackage{amsmath,amssymb,amsthm,latexsym,epsfig,float}

\begin{document}

\begin{frontmatter}

\title{Ultrametrics in the genetic code and the genome}

\author{Branko Dragovich$^{a,b}$, Andrei Yu. Khrennikov$^{c}$, Nata\v sa \v Z. Mi\v si\'c$^{d}$ }

\address{$^a$Institute of Physics, University of Belgrade, Belgrade, Serbia \\
$^b$Mathematical Institute, Serbian Academy of Sciences and Arts, Belgrade, Serbia \\
 $^c$International Center for Mathematical Modeling in Physics, Engineering, Economics, \\
and Cognitive Science, Linnaeus University, S-35195, V\"axj\"o, Sweden \\
$^d$Lola Institute, Kneza Vi\v seslava 70a, Belgrade, Serbia}



\begin{abstract}
  Ultrametric approach to the genetic code and the genome is considered and developed. $p$-Adic degeneracy of the genetic code
  is pointed out. Ultrametric tree of the codon space is presented. It is shown that codons and amino acids can be treated as
  $p$-adic ultrametric networks. Ultrametric modification of the Hamming distance is defined and noted how it can be useful.
 Ultrametric approach with $p$-adic distance is an attractive and promising  trend towards investigation of bioinformation.
\end{abstract}


\end{frontmatter}

\section{Introduction}

 The choice of mathematical methods in the investigation of physical systems depends on their space and time scale as well as of their complexity.
 Sometimes standard methods are not sufficient and one has to invent a new advanced method. Biological systems belong to the most complex
 systems in the nature. In particular, biosystems related to the information processing are very complex and they cannot be completely reduced to the standard
physical systems -- they are something more than ordinary physical systems and need some new theoretical concepts and mathematical methods
to their description and understanding.

It is well known that there is  a strong relation between structure and function in living matter. In bioinformation systems we should
consider not only  physical but also information structure. In the case of physical structure, we use ordinary metrics of Euclidean (or Riemannian) geometry.
It is very important to have a metrics which could appropriately  describe the structure of a bioinformation as well as similarity
(or dissimilarity) between two bioinformation. When we have finite strings (words) of equal length, which are composed of a few
different elements (letters), then usually the Hamming distance is used to measure number of positions at which elements (letters) differ.
Note that dissimilarity is complementary property to similarity, i.e.  less  dissimilarity -- more similarity, and vice versa. So,
one can say that such two strings are more similar as the Hamming distance between them is smaller.
However, Hamming distance is not appropriate when informational content of structure elements depends on their place (hierarchy) in the string,
e.g.  when meaning of elements at the beginning is more important than those at the end. In such case, an ultrametric distance
is just an appropriate tool to measure dissimilarity and then bioinformation system can be regarded as an ultrametric space.

Note that an ultrametric space is a metric space in which distance satisfies strong triangle inequality instead of the ordinary  one, i.e.
 $  d(x,y) \leq \text{max} \{d(x,z),d(z,y) \} .$  As a consequence  of this ultrametric inequality, the ultrametric spaces have some rather
 unusual properties, e.g. all triangles are isosceles with one side  which cannot be larger than the other two.
The Baire metrics  between two different words defined to  be $2^{-m +1}$, where $m$ is the first position at which the words differ,
is an ultrametric distance. Ultrametrics with $p$-adic distances belong to the most elaborated and informative ultrametric spaces.
Ultrametrics has natural application  in the taxonomy, phylogenesis, genetic code and some complex physical systems \cite{Virasoro}. Having many unusual properties,
ultrametrics cannot be represented in the Euclidean space, however it can be illustrated in the form of a tree, dendrogram  or a fractal.

 In this paper we reconsider and further develop $p$-adic approach to the genetic code and the genome introduced in paper \cite{branko1} and
 considered in \cite{branko2,branko3,branko4}. Similar  model of the genetic code was considered on diadic plane \cite{kozyrev}, see also \cite{kozyrev1}.
 A dynamical model of the genetic code origin is presented in \cite{andrei}.
 In Sec. 2 some basic properties of ultrametric spaces are presented and illustrated by a few elementary examples with ordinary, the Baire and $p$-adic metrics.
 Sec. 3 contains the basic notions of molecular biology including DNA, RNA, codons, amino acids and the genetic code. It also contains the  ultrametric trees of
 codons and amino acids. $p$-Adic structure of the genetic code is described in Sec. 4, which also contains the ultrametric network aspects of the genetic code. Some
 $p$-adic ultrametrics
 of the genome is considered in Sec. 5. The last section is devoted to conclusion and  concluding remarks.

\section{Ultrametric spaces}

The general notion of metric space $(M,d)$ was introduced in 1906
by M. Fr\'echet (1878--1973), where $M$ is a set and $d$ is a
distance function. Recall that distance $d$ is a  real-valued function of any
two elements $x,y \in M$ which must satisfy the following
properties:
$(i) \,\, d(x,y) \geq 0, \, \,  d(x,y) = 0 \Leftrightarrow x=y, \quad
(ii) \,\,  d(x,y) = d(y,x), \quad
(iii) \,\,  d(x,y) \leq d(x,z) + d(z,y).
$
Property $(iii)$ is called the triangle inequality. An ultrametric space  is a metric space where the triangle inequality is replaced by
\begin{align}
  \quad  d(x,y) \leq \text{max} \{d(x,z),d(z,y) \},   \label{1.1}
\end{align}
which is called the strong triangle (also ultrametric or non-Archimedean) inequality.
Strong triangle inequality \eqref{1.1} was formulated in 1934 by F. Hausdorff (1868--1942) and ultrametric  space was introduced  by M. Krasner
(1912--1985) in 1944.

As a consequence of the ultrametric inequality \eqref{1.1}, the ultrametric spaces have many unusual properties. It is worth mention some of them.
\begin{itemize}
\item {\it All triangles are isosceles.} This can be easily seen, because any three points $x,y,z$ can be arranged so that inequality \eqref{1.1}
can be rewritten as $d(x,y) \leq d(x,z) = d(z,y) .$
\item {\it There is no partial intersection of the balls.} {\it Any point of a ball can be its center.} {\it Each ball is both open and closed -- clopen ball.}
For a proof of these properties of balls, see e.g. \cite{Schikhof}.
\end{itemize}

\subsection{Simple examples of finite ultrametric spaces}

Without loss of generality, we are going to present  some examples constructed by an alphabet  with fixed
length $n$ of words endowed with an ultrametric distance. Let $m$ ($m =  1, 2, ..., n$) be the first position in a pair of words at which
letters differ counting from their beginning. Thus $m-1$ is the longest common prefix. Then ultrametrics tell us: the longer common prefix,
the closer (more similar) a pair of two words. As  illustrative examples, we will take an alphabet of four letters $\mathcal{A} = \{a, b, c, d\}$
and words of length:   $n= 1, 2, 3$. Let $W_{k,n} (N)$ be a set of words of an alphabet, where $k$ is the number of letters,  $n $ is
the number of letters in words (length of words) and $N$ is the number of words. Then  we have three sets of words:
$ (i)\, W_{4,1}(4); \, \, (ii) \, W_{4,2}(16); \, \, (iii) \, W_{4,3}(64) $  (see Table 1). Note that $N = k^n$. In the following we
will present  ultrametrics of these three different sets with three different distances.
\begin{table} \begin{center} \small{
{\begin{tabular}{|c|c|c|c|}
 \hline
   1 \ \bf a & 2 \ \bf b & 3 \ \bf c & 4 \ \bf d  \\
 \hline
 \hline \  & \   &  \  &   \\
   11\ \bf aa & 21 \bf ba & 31 \bf ca & 41 \bf da  \\
   12\ \bf ab & 22 \bf bb & 32 \bf cb & 42 \bf db  \\
   13\ \bf ac & 23 \bf bc & 33 \bf cc & 43 \bf dc  \\
   14\ \bf ad & 24 \bf bd & 34 \bf cd & 44 \bf dd \\
  \hline
  \hline \  & \     & \ &  \\
  111 \bf aaa & 211 \bf baa & 311 \bf caa & 411 \bf daa  \\
  112 \bf aab & 212 \bf bab & 312 \bf cab & 412 \bf dab  \\
  113 \bf aac & 213 \bf bac & 313 \bf cac & 413 \bf dac  \\
  114 \bf aad & 214 \bf bad & 314 \bf cad & 414 \bf dad  \\
 \hline \  & \  &  \   &   \\
  121 \bf aba & 221 \bf bba & 321 \bf cba & 421 \bf dba  \\
  122 \bf abb & 222 \bf bbb & 322 \bf cbb & 422 \bf dbb  \\
  123 \bf abc & 223 \bf bbc & 323 \bf cbc & 423 \bf dbc  \\
  124 \bf abd & 224 \bf bbd & 324 \bf cbd & 424 \bf dbd \\
 \hline \  & \   & \   &   \\
  131 \bf aca & 231 \bf bca & 331 \bf cca & 431 \bf dca  \\
  132 \bf acb & 232 \bf bcb & 332 \bf ccb & 432 \bf dcb  \\
  133 \bf acc & 233 \bf bcc & 333 \bf ccc & 433 \bf dcc  \\
  134 \bf acd & 234 \bf bcd & 334 \bf ccd & 434 \bf dcd  \\
 \hline \  & \   & \   &   \\
  141 \bf ada & 241 \bf bda & 341 \bf cda & 441 \bf dda  \\
  142 \bf adb & 242 \bf bdb & 342 \bf cdb & 442 \bf ddb  \\
  143 \bf adc & 243 \bf bdc & 343 \bf cdc & 443 \bf ddc  \\
  144 \bf add & 244 \bf bdd & 344 \bf cdd & 444 \bf ddd  \\
\hline
\end{tabular}}{} } \end{center}
\caption{This is table of words constructed of four letters and arranged in the ultrametric form. The same has done with $5$-adic numbers, where four
digits are identified as
$a=1, \, b=2, \, c=3, \, d=4. $  The above three rectangles illustrate ultrametric spaces as follows: $(i) \, W_{4,1}(4)$ at the top; \ $(ii) \, W_{4,2}(16)$
between top and bottom; $(iii) \, W_{4,3}(64)$ at the bottom.
Case $(i) \, W_{4,1}(4):$ Ordinary, Baire and $p$-adic distance are the same and equal 1,
when prime $p \geq 5.$ However there are examples when $p$-adic distance is smaller than 1, i.e. $d_2 (3,1) = d_2 (4,2) = \frac{1}{2}$ and
$d_3 (4,1) = \frac{1}{3} .$
Case $(ii) \, W_{4,2}(16):$ Note that in the columns, the first digits (letters) are the same and otherwise distinct, what expresses the ultrametric
similarity and dissimilarity, respectively.
Case $(iii) \, W_{4,3}(64):$ Here $64$ three-digit $5$-adic numbers (three-letter words) are presented so
that within boxes $5$-adic distance is the smallest, i.e. $d_5(x, y) = \frac{1}{25}, $ while  $5$-adic distance between any two boxes in
vertical line is $\frac{1}{5}$ and otherwise is equal $1.$  Ultrametric tree illustration of these three cases is in Fig. 1.
\label{Tab:1}}
\end{table}

{\bf Ordinary ultrametric distance.} Let us define ordinary ultrametric  distance between any two different words $x$ and $y$  as $d(x,y) = n - (m-1) .$  It takes   $n$ values, i.e. $d(x,y) = 1, 2, ..., n$. Note that one can redefine this distance by scaling it as  $d_s(x,y ) = \frac{n - m+1}{n} $ and then the scaled distances  are between $1$ and $\frac{1}{n}$.
\begin{itemize}
\item {\it $(i)\, Case \, W_{4,1}(4).$} In this case letters $a, b, c, d, $ are words as well. The distance between any two words (letters) is $1$, because  $n = 1$ and $m = 1.$

\item {\it  $(ii)\, Case \, W_{4,2}(16).$} Here we have two-letter words (see Table 1). The distance between any two different words $x$ and $y$ is
$d(x, y) = 2$ when letters differ at the first position and $d(x, y) = 1$ if letters at the first position are the same  $(m = 2)$. Scaling distance is
   \begin{equation}
    d_s(x,y ) = \frac{2 - m+1}{2} = \begin{cases} 1, \, & m = 1 \\ \frac{1}{2}, \, &m = 2 . \end{cases} \label{2.2}
  \end{equation}

\item {\it $(iii)\, Case \, W_{4,3}(64).$} Now we have three-letter words (see Table 1). Possible values of distance $d (x,y)$ are $1, 2, 3.$ the corresponding scaling distance is
    \begin{equation}
    d_s(x,y ) = \frac{3 - m+1}{3} = \begin{cases} 1, \, & m = 1 \\ \frac{2}{3}, \, &m = 2 \\ \frac{1}{3}, \, &m = 3 . \end{cases} \label{2.2}
  \end{equation}

\end{itemize}

{\bf The Baire distance.}  This distance can be defined as $d_B (x, y) = 2^{-(m-1)} ,$ where $m$ is as defined in the above, i.e. it is the first position in words $x$ and $y$ at which letters differ, i.e. $m = 1, 2, ..., n.$ Thus the Baire distance takes values $1, \frac{1}{2}, \frac{1}{2^2}, ..., \frac{1}{2^{n-1}}.$  Note that instead of the base $2$  one can take any integer larger than $2$.
\begin{itemize}
\item {\it  $(i) \, Case \, W_{4,1}(4).$}  Now $d_B (x, y) = 1 ,$ i.e. the same as in the ordinary ultrametric case.

\item {\it  $(ii) \, Case \, W_{4,2}(16).$}
    \begin{equation}
    d_B(x,y ) = 2^{-(m-1)} = \begin{cases} 1, \, & m = 1 \\ \frac{1}{2}, \, &m = 2 . \end{cases} \label{2.2}
  \end{equation}

\item {\it $(iii) \, Case \, W_{4,3}(64).$} In this case the Baire distance is
 \begin{equation}
    d_B(x,y ) = 2^{-(m-1)}  = \begin{cases} 1, \, & m = 1 \\ \frac{1}{2}, \, &m = 2 \\ \frac{1}{4}, \, &m = 3 . \end{cases} \label{2.2}
  \end{equation}

\end{itemize}

{\bf $p$-Adic distance.} Recall that $p$-adic norm ($p$-adic absolute value) of an integer $x$ is $|x|_p = p^{-k} ,$ where $k$ is
degree of a prime number $p$ in $x$. Since $k = 0, 1, 2, ... ,$  $p$-adic norm of any integer $x$ is $|x|_p \leq 1 .$ By definition,
$p$-adic distance between two  integers $x$ and $y$ is $d_p (x, y) = |x-y|_p ,$  i.e. this distance is related to divisibility of $x-y$
by prime $p$  (more divisible - lesser distance).  Recall also that any integer, with respect to a fixed prime $p$ as a base, can be
expanded in the unique way, e.g. $x = x_0 + x_1 \, p + x_2 \, p^2 + ... + x_n \, p^n ,$ where $x_i \in \{ 0, 1, ..., p-1 \}$ are digits.
If $x_k $ is the first digit different from zero,  then $p$-adic norm of this $x$ is $|x|_p = p^{-k} .$  To have connection with the
above alphabet and words  it is natural to make a correspondence between letters and  digits, e.g. by identification of four letters
$\{a, b, c, d\}$ with four digits $\{x_0, \, x_1, \, x_2, \, x_3 \} .$ In this way the role of letters play digits (see Tab. 1).
The smallest prime number which can be used as base and contains four digits is $p =5$ and we will use digits $\{1,\,2,\, 3, \, 4\}$
without digit $0$. Skipping  digit $0$ is suitable in $p$-adic modeling of the genetic code.
Namely, to use the digit $0$ for a nucleotide is inadequate, because it may lead to non-uniqueness in the representation
of the codons by natural numbers in DNA and RNA. For example, if we use digit $0$ for a nucleotide then $121000$ denotes sequence of two
codons (121 and 000), but the corresponding natural number is the same for $121000$ and $121$ (see notation below).
  Hence we will use some sets of  $5$-adic integers in the form:
\begin{equation}
x = x_0 + x_1 \, 5 + ...+ x_k \, 5^k  \quad  \text{or} \quad x \equiv x_0 x_1 ... x_k, \quad x_i \in \{1, 2, 3, 4\} . \label{2.1}
\end{equation}

\begin{itemize}
\item {\it $(i) \, Case \, W_{4,1}(4).$} In this simplest case $x = x_0 ,$ where $x_0 \in \{1, 2, 3, 4\}$. The corresponding $5$-adic distance between different words (digits) $x = x_0$ and $y =y_0$ is $d_5 (x, y) = |x_0 - y_0|_5 = 1 .$

\item {\it  $(ii) \, Case \, W_{4,2}(16).$} Now we have $16$ numbers (words) in the form $ x = x_0 + x_1 5 .$  The $5$-adic distance between numbers $x = x_0 + x_1 5$ and $y = y_0 + y_1 5$ is
    \begin{equation}
    d_5 (x,y) = |x_0 + x_1 5 - y_0 - y_1 5|_5 = \begin{cases} 1, \, &x_0 \neq y_0 \\ \frac{1}{5}, \, &x_0 =y_0, x_1 \neq y_1 . \end{cases} \label{2.2}
  \end{equation}

\item {\it  $(iii) \, Case \, W_{4,3}(64).$} In this case we have three-letter words represented by three-digit $5$-adic numbers (see Table 1). The corresponding $5$-adic distance of a pair of words (numbers) $x = x_0 + x_1 5 + x_2 5^2 \equiv x_0 x_1 x_2$ and $y = y_0 + y_1 5 + y_2 5^2 \equiv y_0 y_1 y_2$ is:
    \begin{equation}
d_5 (x,y) = |x_0 x_1  x_2  - y_0  y_1  y_2 |_5 = \begin{cases} 1, \, &x_0\neq y_0 \\ \frac{1}{5},  \, &x_0 = y_0, x_1\neq y_1  \\ \frac{1}{25},  \, &x_0 = y_0, x_1= y_1, x_2 \neq y_2 \, .  \end{cases}  \label{2.3}
\end{equation}

\end{itemize}

Note that $p$-adic distance between words is finer and  more informative than the ordinary and the Baire distances.
Namely, for the same set of  natural numbers one can also employ $p$-adic distance with $p \neq 5 .$ For example, in the $p$-adic case $W_{4,1}(4)$ we have
$d_2 (1,3) = d_2 (2,4) = |2|_2 = \frac{1}{2}$ and $d_3 (1,4) = |3|_3 =\frac{1}{3}$, while other $2$-adic and $3$-adic distances are equal to $1$.

In fact, the most advanced example of the ultrametric spaces is the field of $p$-adic numbers $\mathbb{Q}_p .$  $p$-Adic numbers are discovered by
K. Hensel (1861--1941) in 1897. Many of their mathematical  aspects have been elaborated, see e.g. books \cite{Gelfand,Schikhof}. Many applications from
Planck scale physics via complex systems to the universe as a whole, known as
$p$-adic mathematical physics, have been considered, e.g. see \cite{VVZ,Khrennikov1,Khrennikov2} as books, \cite{Freund,Dragovich} as review articles,
and \cite{AIP,journal} as conference proceedings and related journal. $p$-Adic mathematical physics has inspired investigations in some other fields like
nonlocal modified gravity, see e.g. \cite{Dimitrijevic}.

From the above examples we see that the ultrametric distance directly measures dissimilarity between two words, or in other words,
dissimilarity between two elements of an ultrametric space.

All the above ultrametric examples can be represented as  trees. Namely, instead of letters $\{a, b, c, d\}$ or digits $\{1, 2, 3, 4\}$ one
can take four  line segments (of different colors) to draw edges of the related tree   (see Fig. 1).

\begin{figure}[t]
\begin{center}
\includegraphics[width=11cm]{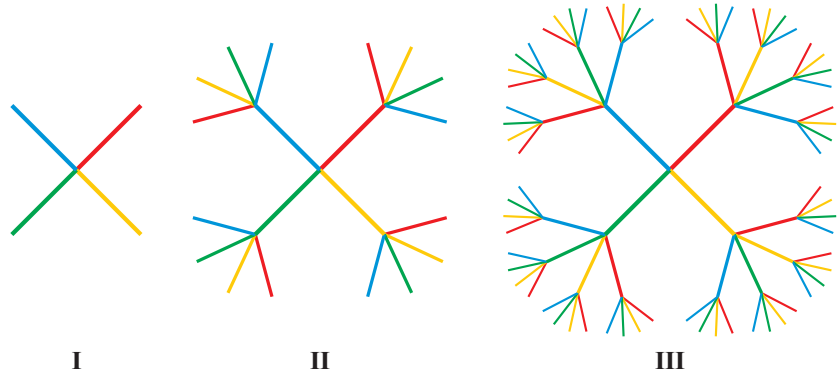}
\caption{Utrametric trees related to Table 1. Tree I, II and III are related to top ($W_{4,1}(4)$), intermediate ($W_{4,2}(16)$)
and bottom ($W_{4,3}(64)$) case, respectively. Ultrametric tree III is also related to the vertebrate mitochondrial code presented
at the Table 2. One can easily calculate ordinary ultrametric distance
and see that distance between any three tree end points satisfies the strong triangle (ultrametric) inequality.}
\end{center}
\end{figure}

\section{Ultrametric approach to the genetic code}

To show comprehensively that the genetic code has an ultrametric structure, and in particular $p$-adic structure, it is worth first to recall some  relevant notions from molecular biology.

\subsection{Some notions of molecular biology}

{\bf DNA and RNA.}  Genotype information is stored in the DNA (deoxyribonucleic acid)  which  is a macromolecule composed
of two polynucleotide chains with a double-helical structure.
The building blocks of genetic information are nucleotides which  consist of a base, a
sugar and a phosphate group. There are  four bases
called: cytosine ($C$), adenine ($A$),  thymine ($T$) and guanine ($G$).   Cytosine and thymine
are pyrimidines and they have one carbon-nitrogen ring with two nitrogen atoms.
Adenine and guanine are  purines, which contain two carbon-nitrogen rings with four nitrogen atoms.
In the sense of information, the nucleotide and
its base  have the same meaning. Nucleotides are arranged along
chains of double helix through base pairs $A-T$ and $C-G$ bonded by two
and three hydrogen bonds, respectively. As a consequence of this
pairing there is in the DNA an equal number of cytosine and guanine, as well
as the equal number of adenine and thymine. DNA is packaged in
chromosomes, which  in the eukaryotic cells  are localized in the nucleus.
Human genome, as complete genetic information in human cell,  is contained in 23 chromosome pairs and  mitochondria, with  about 3 billion DNA base pairs.
Only about $1.5 \% $ of DNA is protein-coding part,
while the rest is partially related to some regulation processes.


By
transcription,  a gene in DNA is transcribed by synthesis of  the messenger ribonucleic
acid (mRNA), which is usually a single-stranded  polynucleotide chain. During synthesis of mRNA  nucleotides $C, A, T, G$ from DNA are
respectively transcribed into their complements $G, U, A, C,$ where $T$ is
replaced by $U$ ($U$ is the uracil, which is a
pyrimidine). The next step in gene expression is translation, when
the information coded by codons in the mRNA  is translated into amino acids, which are building blocks in
synthesis of proteins.

{\bf Codons and amino acids.} Codons are ordered  trinucleotides composed of $C, A, U (T)$ and $G$. There are $64$ codons.
Each of them is an information which strictly determines  one of
the 20 standard amino acids or stop signal in synthesis of
proteins.

The whole complex process of protein synthesis is carried out by the ribosome.
Proteins
are organic macromolecules composed of amino acids arranged in a
linear chain, which in the process of folding gets a definite spatial structure.
They are  the most diverse biomolecules on our planet and substantial ingredients of all living
organisms participating in various processes in cells and
determine the phenotype of an organism  \cite{Finkelshtein}.

\begin{table} \begin{center}
  \small{ {\begin{tabular}{|l|l|l|l|}
 \hline \ & \ & \ & \\
  111 \, CCC \, Pro &   211 \, ACC \, Thr  &  311 \, UCC \, Ser &  411 \, GCC \, Ala  \\
  112 \, CCA \, Pro &   212 \, ACA \, Thr  &  312 \, UCA \, Ser &  412 \, GCA \, Ala  \\
  113 \, CCU \, Pro &   213 \, ACU \, Thr  &  313 \, UCU \, Ser &  413 \, GCU \, Ala  \\
  114 \, CCG \, Pro &   214 \, ACG \, Thr  &  314 \, UCG \, Ser &  414 \, GCG \, Ala  \\
 \hline \  & \  &  \ & \ \\
  121 \, CAC \, His &   221 \, AAC \, Asn  &  321 \, UAC \, Tyr &  421 \, GAC \, Asp  \\
  122 \, CAA \, Gln &   222 \, AAA \, Lys  &  322 \, UAA \, Ter &  422 \, GAA \, Glu  \\
  123 \, CAU \, His &   223 \, AAU \, Asn  &  323 \, UAU \, Tyr &  423 \, GAU \, Asp  \\
  124 \, CAG \, Gln &   224 \, AAG \, Lys  &  324 \, UAG \, Ter &  424 \, GAG \, Glu  \\
 \hline \  & \  & \  &   \\
  131 \, CUC \, Leu &   231 \, AUC \, Ile  &  331 \, UUC \, Phe &  431 \, GUC \, Val  \\
  132 \, CUA \, Leu &   232 \, AUA \, Met  &  332 \, UUA \, Leu &  432 \, GUA \, Val  \\
  133 \, CUU \, Leu &   233 \, AUU \, Ile  &  333 \, UUU \, Phe &  433 \, GUU \, Val  \\
  134 \, CUG \, Leu &   234 \, AUG \, Met  &  334 \, UUG \, Leu &  434 \, GUG \, Val  \\
 \hline \ & \   & \  &   \\
  141 \, CGC \, Arg &   241 \, AGC \, Ser  &  341 \, UGC \, Cys &  441 \, GGC \, Gly  \\
  142 \, CGA \, Arg &   242 \, AGA \, Ter  &  342 \, UGA \, Trp &  442 \, GGA \, Gly  \\
  143 \, CGU \, Arg &   243 \, AGU \, Ser  &  343 \, UGU \, Cys &  443 \, GGU \, Gly  \\
  144 \, CGG \, Arg &   244 \, AGG \, Ter  &  344 \, UGG \, Trp &  444 \, GGG \, Gly  \\
\hline
\end{tabular}}{}} \end{center}
\caption{ The vertebrate mitochondrial code with $p$-adic ultrametric structure. Digits are related to nucleotides as follows: $C=1,\, A=2,\, U=3,\, G=4$. $5$-Adic distance between codons:  $\frac{1}{25}$ inside quadruplets,
 $\frac{1}{5}$ between different quadruplets in the same column, $1$ otherwise.  Each quadruplet  can be viewed as two doublets, where every doublet code
 one amino acid or termination signal (Ter). $2$-Adic distance  between codons in doublets is $\frac{1}{2}$. Two doublets which code the same aa belong
 to the same quadruplet. Amino acids leucine (Leu) and serine (Ser) are coded by three doublets  -- the third doublet is at $\frac{1}{2}$ $2$-adic distance
 with respect to the corresponding doublet in quadruplet, which contains the first two doublets.  \label{Tab:2} }
\end{table}

Amino acids are molecules that consist of the  amino,
carboxyl and R (side chain) groups. Depending on R group there are
20 standard amino acids (aa). These amino acids are joined together by
a peptide bond. The sequence of amino acids in a protein is determined by ordered sequence
of codons contained in genes.  The informational  connection between codons and
amino acids with stop signal is known as the  genetic code (GC).

{\bf The genetic code.}
From mathematical point of view,  the GC is a map from a set of $64$ elements onto a set of $21$ element.
There is a huge number of possible such maps. Namely,
if each amino acid and stop signal are coded by at least one codon, then the total number of  possible maps
is more than $10^{84}$ \cite{Koonin}.  However, it is presently
known only a few dozens of codes in living organisms. The most important
are two of them: the standard code and the  vertebrate
mitochondrial code. We shall mainly consider this mitochondrial code and all other codes can be
viewed as its slight modification. It is worth noting that all known  codes have many common
characteristics, e.g. four nucleotides, trinucleotide
codons, the similar procedure of protein synthesis and many others.

After discovery of DNA structure by Crick and Watson in 1953, there have been many papers devoted to theoretical modeling of the genetic code.
For a popular review of the early models, see \cite{Hayes}.  The genetic code has many aspects which caused its investigation from many points of view --
mathematical, physical, chemical, biological and others, see e.g. \cite{Koonin,Rumer,Crick,Wong,Swanson,Osawa,Hornos,Forger,Shcherbak,Rakocevic,Misic}
and references therein. Nevertheless, there is not yet a complete description and understanding of the genetic code.  In this paper we further develop
$p$-adic model to the genetic code, introduced in \cite{branko1}, and push forward ultrametric approach to bioinformation.

In the case of the vertebrate mitochondrial code (VMC), $64$ codons can be viewed as $32$ codon doublets,
which are distributed as follows: 12 amino acids (His, Gln, Asn, Lys, Tyr, Asp, Glu, Ile, Met, Phe, Cys and Trp) are coded by single doublets,
6 aa (Pro, Thr, Ala, Val, Arg and Gly) and stop signal are related to two doublets, and
2 aa (Ser and Leu) are coded by three doublets. Thus we see that each of amino acids is coded either by two, four or six codons.
The property that some (in this case all) aa are coded by more than one codon  is called {\it degeneracy} of the GC. In principle,
the degeneracy could be emerged in a very large number of ways,
but life on the Earth was evolved by only a few of them.  It is obvious  that code degeneracy of the vertebrate mitochondria is
not random but highly regular (see Tab.2). We  show that the GC degeneracy has $p$-adic ultrametric structure.
The ultrametric degeneracy is a very useful property, because it  minimizes errors caused by mutations.

Note that the standard GC can be obtained from this mitochondrial one by the following formal replacements in codon assignments:
AUA: Met $\to$ Ile; AGA and AGG: Ter $\to$ Arg; UGA:
Trp $\to$ Ter.

From linguistic point of view, the GC is a dictionary that translates one language of four letters (nucleotides) into
another  language of twenty letters (amino acids). On the one hand, there are 64 three-letter words called codons, and on
the other one there are thousands many-letter words known as proteins. These are two natural biomolecular languages inside
cells -- at the first language life is coded from generation to generation and at the second one life mainly functions.

{\bf Ultrametric tree of the codon space and amino acids.}
  We want first to point out ultrametric structure of the codon space and  in a sense of amino acids.
  Then we will use $p$-adic distance to describe ultrametrics of the genetic code.

  The vertebrate mitochondrial code is presented at Tab. 2.
 Comparing this VMC  and  {\it $(iii)\, Case \, W_{4,3}(64)$} at  Tab. 1  one can easily observe similarity and conclude
 that $64$ codons are arranged      in the same ultrametric way. Moreover, we can identify two alphabets: $C = a, \ A = b, \ U = c, G = d .$
 It is obvious that $64$ codons  make an ultrametric space and can be illustrated  in the form of ultrametric tree presented at Fig. 1, III.

  Is there any ultrametric structure in the set of $20$ amino acids?  From the point of view of the genetic code, answer to this question is positive.
  Namely, there are $8$ codon quadruplets which  code $8$ amino acids (Pro, Thr, Ser, Ala, Leu, Val, Arg, Gly) practically by first two nucleotides, because result
  does not depend on the third nucleotide (see Tab. 2). There are  additional $8$ codon doublets with cytosine (C) or uracil (U) at the third position, which
  code $8$ amino acids (His, Asn, Tyr, Asp, Ile, Phe, Ser, Cys) and practically also by first two nucleotides. Another $8$ doublets (with adenine (A)
  or guanine (G) at the third position) are unstable in their coding amino acids and stop signal, and lead to other versions of the genetic code. By this reasoning
  we conclude that there are $16$ dinucleotides  which firmly  code $15$ amino acids (because Ser is coded twice) to which one can attach two nucleotide letters and
  two $5$-adic digits. This is presented at Tab. 3 (see also intermediate box at Tab. 1 and Fig. 1, II).


\begin{table}
 \begin{center}
 {\begin{tabular}{|c|c|c|c|}
 \hline \ & \ & \ & \\
 11(11)  CC \, Pro & 21(12)  AC \, Thr  & 31(13)  UC \, Ser & 41(14)   GC \, Ala  \\
  \hline \  & \  &  \ & \ \\
 12(21)  CA \, His &  22(22)  AA \, Asn  & 32(23)  UA \, Tyr & 42(24)  GA \, Asp  \\
  \hline \  & \  & \  &   \\
 13(31)  CU \, Leu &  23(32)  AU \, Ile  & 33(33)  UU \, Phe & 43(34)  GU \, Val \\
  \hline \ & \   & \  &   \\
 14(41)  CG \, Arg &  24(42)  AG \, Ser  & 34(43)  UG \, Cys & 44(44)  GG \, Gly  \\
 \hline
\end{tabular}}{}
\caption{ Table of amino acids coded by the codons which have pyrimidine at the third position. Only serine (Ser) appears twice. By this way,
 there is a formal connection between the amino acids and the root (dinucleotide) of codons coding them. Identifying these amino acids with
 related codon roots (i.e. first two digits of $5$-adic numbers) one gets some ultrametricity between above amino acids (on importance of $16$
  codon roots, see \cite{Rumer}).
 Since the amino acids which are coded by codons having the same nucleotide at the second position have the similar chemical properties, it
 is better to use ultrametric distance assigning digits to  amino acids in opposite way, as it is done in the brackets. This interchange of
 digits could be related to evolution of the genetic code \cite{branko2}. } 
\end{center}
\end{table}

\section{The $p$-adic  genetic code}

Ultrametric structure of the codon space demonstrated above can be described by $5$-adic and $2$-adic distance in the more concrete form.

{\bf $5$-Adic and $2$-adic structure of the codon space.}
The first question we have to analyze here is related to  the most adequate connection between the set of
nucleotides $\{ C, A, U, G\}$ and the set of digits $\{ 1, 2, 3, 4\}.$ From the first sight it follows that there are 4! possibilities.
However, taking into account the chemical properties of nucleotides and coded amino acids, 24 possibilities
can be reduced to 8 options presented at Tab. 4. Namely, on the one side there are two pyrimidines which have similar
structure (one ring) and coding function. On the other side, there are two purines which also have similar structure (two rings)
and coding function. Fortunately, this similarity within two  pyrimidines, as well as  similarity between two purines, can be
described by $2$-adic distance. Also by $2$-adic distance one can express dissimilarity between purines and pyrimidines.
Since  $d_2 (3,1) = d_2 (4,2) = |2|_2 = \frac{1}{2}$ one has to connect nucleotides and digits so that $d_2 (U,C) = d_2 (G,A) = \frac{1}{2}$ and
$d_2 (purine, pyrimidine) = 1.$ There are $8$ possibilities which satisfy this condition and they are presented at Tab. 4. At Tab. 2 we presented case
$C = 1, A=2, U =3, G = 4.$  If we fix digits with boxes at Tab. 2 and change the connection between digits and nucleotides, then the codon quadruplets
will change their boxes. However, this is not important and we use $C = 1, A=2, U =3, G = 4.$ Moreover, in this case $d_3(4,1) = \frac{1}{3}$ and $d_3(3,2) = 1 ,$ what
could be related to hydrogen bonds of pairs $C-G$ and $A-T$ in DNA, respectively. Note that there is symmetry in distribution of codon doublets and quadruplets
with respect to the middle vertical line at Tab. 2.
\begin{table}
\begin{center}
\small{ {\begin{tabular}{|l|c|c|r|}
 \hline
  C = 1 \qquad   A = 2  \qquad  U = 3 \qquad  G = 4  \\
  U = 1  \qquad   G = 2  \qquad  C = 3 \qquad  A = 4  \\
 \hline
  C = 1 \qquad   G = 2  \qquad  U = 3 \qquad  A = 4  \\
  U = 1 \qquad   A = 2 \qquad  C = 3 \qquad  G = 4  \\
 \hline
  A = 1 \qquad   C = 2  \qquad  G = 3 \qquad  U = 4  \\
  G = 1 \qquad   U = 2  \qquad  A = 3 \qquad  C = 4  \\
 \hline
  A = 1 \qquad   U = 2  \qquad  G = 3 \qquad  C = 4 \\
  G = 1 \qquad   C = 2  \qquad  A = 3 \qquad  U = 4  \\
\hline
\end{tabular}}{} }
\caption{ Eight possible connections between the nucleotides $\{ C, A, U, G\}$ and the digits $\{ 1, 2, 3, 4\}$ which take care that $2$-adic distance
between two pyrimidines (C,U), as well as between two purines (A, G),  is $\frac{1}{2}$.  In Tab. 2 we employ  connection presented in the first row.}
\end{center}
\end{table}

As we mentioned, an amino acid in the VMC  is coded either by one, two or three  pairs of codons. Every such pair of codons has the same first two nucleotides
and at the third position two pyrimidines or two  purines. A pair of two codons  which are simultaneously
at $\frac{1}{25} $ $ 5$-adic distance and $\frac{1}{2} $ $ 2$-adic distance is called codon doublet. There are $32$ codon doublets, such that every of $30$ doublets codes one of $20$ amino acids and $2$ doublets contain stop codons.


\subsection{The genetic code as an ultrametric network}

Many systems have the form of networks, which are the sets of nodes (vertices) joined  together
by links (edges). Examples mainly come from  biological and social  systems. According to the above consideration one can look at the genetic code
as a $p$-adic ultrametric network.

Namely, we can start from two separate systems of biomolecules -- one related to $4$ nucleotides and another based on $20$ standard amino acids.
Four types of nucleotides are chemically linked to a large number of various sequences, which are known as DNA and RNA. Standard amino acids are also
chemically linked and form various peptides and proteins. By the genetic code, amino acids are linked to codons which are the elements of an ultrametric space. Since
standard amino acids can be also formally regarded as the elements of an ultrametric space, one can say that the genetic code links two ultrametric networks to
one larger ultrametric network of $85$ elements (64 codons + 20  aa + 1 stop signal). Note that one can also consider the ultrametric distance between codons
and amino acids with stop signal.

Looking at codons as an ultrametric network with information content,  then they are the  nodes mutually linked by similarity according to  $p$-adic distance.
Recall that there are three possibilities of $5$-adic  distance between codons:  $\frac{1}{25}, \, \frac{1}{5}$ and  $1 .$  With respect to these
distances, we can respectively call the corresponding subsets of codons as small, intermediate and large community. Thus, any codon has 3 neighbors at
distance $\frac{1}{25}$ and makes a small community. Any codon is also linked to 12 and 48 other codons to make an intermediate and large community,
respectively. Hence, any codon belongs simultaneously to a small, intermediate and large community.

\begin{table}
 \begin{center}
 {\begin{tabular}{|l|l|l|l|l|l|l|}
 \hline \ & \ & \ & \ & \ & \ &\\
 11    Pro & 12   Thr  & 13    Ser & 14    Ala &  &  & \\
  \hline \  & \  &  \ & \ & \  &  \ & \ \\
 21    His &  22  Asn  & 23    Tyr & 24    Asp  & 212 Gln  & 222 Lys   & 242 Glu  \\
  \hline \  & \  & \  & \ & \  &  \ & \ \\
 31   Leu &  32   Ile  & 33    Phe & 34    Val & 322 Met   &    &\\
  \hline \ & \   & \  & \ & \  &  \ & \ \\
 41    Arg &    & 43    Cys & 44    Gly & 432 Trp &  & \\
 \hline
\end{tabular}}{}
\caption{ The rewritten and extended Table 3, where the first two digits are replaced. Third digits are added to the amino acids
which are coded by one doublet with purine at the third position. Table contains ultrametrics between amino acids, which corresponds
to some their physicochemical properties. $5$-Adic distance between  amino acids in rows is either
 $\frac{1}{5}$  or $\frac{1}{25}$, otherwise it is equal to $1$.      } 
\end{center}
\end{table}

Physicochemical similarities of amino acids in Tab. 5 are as follows.
\begin{itemize}
\item First row: small size and moderate in hydropathy.

\item Second row: average size and hydrophilic.

\item Third row: average size and hydrophobic

\item Fourth row:  special case of diversity.
\end{itemize}

\section{On $p$-adic ultrametrics in the genome}

In previous section we demonstrated that codons and amino acids are elements of some $p$-adic ultrametric spaces.
Ultrametric approach should be useful also in investigation of similarity (dissimilarity) between definite sequences of DNA,
RNA and proteins. These sequences can be genes, microRNA, peptides, or some other polymers. Since elements of genes (proteins)
are codons (amino acids), which have ultrametric properties, it is natural to use their ultrametric similarity in determination
of  similarity between genes (proteins). It means that one can consider not only ultrametric similarity between two sequences (strings)
but also ultrametrically improved Hamming distance.

\subsection{$p$-Adic modification of the Hamming distance}

Let $a = a_1\, a_2\, \cdots a_n$ and $b = b_1\, b_2\, \cdots b_n$
be two strings of equal length. Hamming distance between these two
strings is $d_H (a,b) = \sum_{i=1}^n d(a_i, b_i), $ where $d(a_i,
b_i) = 0$ if $a_i = b_i,$ and $d(a_i, b_i)= 1$ if $a_i \neq b_i .$
In other words, $d_H (a,b) = n -\nu ,$ where $\nu$ is the number of positions at
which elements of both strings are equal.
We introduce $p$-adic  Hamming distance in the
following way: $d_{pH} (a,b) = \sum_{i=1}^n d_p(a_i, b_i), $ where
$d_p(a_i, b_i) = |a_i - b_i|_p$ is $p$-adic
 distance between  numbers $a_i$ and $b_i .$ When $a_i, b_i \in \mathbb{N}$
then $d_p(a_i, b_i) \leq 1 .$ If also $a_i - b_i \neq 0$ is
divisible by $p$ then $d_p(a_i, b_i) < 1.$ There is the following relation:
$d_{pH} (a,b) \leq d_{H} (a,b) \leq d (a,b) ,$ where $d (a,b)$ is ordinary ultrametric distance. In the case of strings
as parts of DNA, RNA and proteins, this modified distance is finer
and should be more appropriate than Hamming distance itself. For
example, elements $a_i$ and $b_i$ can be nucleotides, codons and
amino acids with above assigned natural numbers, and primes $p=2$
and $p=5$.

To illustrate an advantage of the $p$-adic modified Hamming distance with respect to the ordinary Hamming one, it is worth to consider comparison of two sequences whose elements are codons. For simplicity, let  sequences have the three codons.
\begin{itemize}
\item {\em Case (i).} If  $a = a_1\, a_2\, a_3 = 111412443$ and $b = b_1\, b_2\, b_3 = 113414441 $  then the corresponding Hamming distance is
$d_H (a, b) = 3$, while $p$-adic modified ones are $d_{5H} (a, b) = \frac{3}{25}$  and $d_{2H} (a, b) = \frac{3}{2} .$
Now suppose that we do not know exactly these two sequences
$a$ and $b$, but we have information on their distances. If we would know only the Hamming distance we could not  conclude at which  three positions  of related codons nucleotides differ.
However, taking $5$-adic and $2$-adic modified Hamming distances together, it follows that codon differences are at the third position of nucleotides and that sequences $a$ and $b$
code the same sequence of amino acids, in fact the sequence  $ProAlaGlu .$

\item {\em Case (ii).} Let $\bar{a} = \bar{a}_1\, \bar{a}_2\, \bar{a}_3 = 111124434$ and $\bar{b} = \bar{b}_1\, \bar{b}_2\, \bar{b}_3 = 131144414 .$  Then
$d_H (\bar{a}, \bar{b}) = 3$, $d_{5H} (\bar{a}, \bar{b}) = \frac{3}{5}$  and $d_{2H} (\bar{a}, \bar{b}) = \frac{3}{2} .$ From $d_{5H}$ follows that codon counterparts in the  sequences $\bar{a}$ and $\bar{b}$ have the same first nucleotides.

\item {\em Case (iii).} Let $\tilde{a} = \tilde{a}_1\, \tilde{a}_2\, \tilde{a}_3 = 111241344$ and $\tilde{b} = \tilde{b}_1\, \tilde{b}_2\, \tilde{b}_3 = 311441144 .$  Then
$d_H (\tilde{a}, \tilde{b}) = 3$, $d_{5H} (\tilde{a}, \tilde{b}) = 3$  and $d_{2H} (\tilde{a}, \tilde{b}) = \frac{3}{2} .$ In this case one can conclude that
sequences $\tilde{a}$ and $\tilde{b}$ differ at the first nucleotide positions of the related codons.

\end{itemize}
Note that cases $(ii)$ and $(iii)$ are obtained by cyclic permutations of nucleotides inside codons of the case $(i)$. In all three cases  $d_H =3$
 and $d_{2H} = \frac{3}{2},$ but $d_{5H}$ distance is $\frac{3}{25}$, $\frac{3}{5}$ and $3$, respectively. From $d_{2H}$ distances, one can conclude that in the above cases the corresponding nucleotides in related codons are either  purines  or pyrimidines. Unlike to $d_{5H}$ and $d_{2H}$, the ordinary Hamming distance
tell us only that there is a distinction between the corresponding codons.

\section{Concluding remarks}

In this paper we presented three simple examples of ultrametric spaces which are applied to the $p$-adic modeling of $64$ codons and $20$ standard amino acids.
Ultrametric space of codons is illustrated by the corresponding tree. Sixteen dinucleotide codons are also presented with their ultrametric structure by tree
and corresponding table. We emphasize that degeneracy of the vertebrate mitochondrial code has strong ultrametric structure. It is shown that codons
and amino acids can be viewed as ultrametric networks which are connected by the genetic code.
The $p$-adic Hamming distance is defined. Investigation of similarity (dissimilarity) between genes, microRNA, proteins and some other polymers by $p$-adic
ultrametric approach is proposed.

It is worth emphasizing
that our $5$-adic approach, extended by $2$-adic distance, correctly describes mathematical structure of the vertebrate mitochondrial code,
and is in agreement with its chemical and biological aspects.

We plan to employ this ultrametric approach to investigation of concrete DNA, RNA and protein sequences. This approach can be also applied to analyze
similarity of words in some human languages and systems of hierarchical structure. An interesting subject which deserves further investigation
is ultrametric approach to the evolution of
genetic code, see \cite{branko2} and \cite{avetisov}.  Application of $p$-adic ultrametricity to cognitive neuroscience is a big challenge \cite{Khrennikov2,Khrennikov3}.

\section*{Acknowledgments}
This work was supported in part by Ministry of Education, Science and Technological Development of the Republic of Serbia, projects:
OI 173052,  OI 174012, TR 32040 and TR 35023.  Two authors of the paper (B.D. and  A.K.) were partially supported by
the grant Mathematical Modeling of Complex Hierarchic Systems of Linnaeus University.
B.D. also thanks M. Rako\v cevi\'c for useful discussions on chemical aspects of the genetic code. Authors are grateful to the referees for their comments
to improve presentation of this paper.

\end{document}